# Use of Interactive Simulations in Fundamentals of Biochemistry, a LibreText Online Educational Resource, to Promote Understanding of Dynamic Reactions

Henry V. Jakubowski, Henry Agnew, Bartholomew Jardine, and Herbert M. Sauro

**Abstract**

Biology is perhaps the most complex of the sciences, given the incredible variety of chemical species that are interconnected in spatial and temporal pathways that are daunting to understand. Their interconnections lead to emergent properties such as memory, consciousness, and recognition of self and non-self. To understand how these interconnected reactions lead to cellular life characterized by activation, inhibition, regulation, homeostasis, and adaptation, computational analyses and simulations are essential, a fact recognized by the biological communities. At the same time, students struggle to understand and apply binding and kinetic analyses for the simplest reactions such as the irreversible first-order conversion of a single reactant to a product. This likely results from cognitive difficulties in combining structural, chemical, mathematical, and textual descriptions of binding and catalytic reactions. To help students better understand dynamic reactions and their analyses, we have introduced two kinds of interactive graphs and simulations into the online educational resource, Fundamentals of Biochemistry, a multivolume biochemistry textbook that is part of the LibreText collection. One type is available for simple binding and kinetic reactions. The other displays progress curves (concentrations vs time) for both simple reactions and more complex metabolic and signal transduction pathways, including those available through databases using systems biology markup language (SBML) files. Users can move sliders to change dissociation and kinetic constants as well as initial concentrations and see instantaneous changes in the graphs. They can also export data into a spreadsheet for further processing, such as producing derivative Lineweaver-Burk and traditional Michaelis-Menten graphs of initial velocity ($v_0$) vs substrate concentration.

It should come as no surprise to instructors that students struggle with kinetics, either in the absence or presence of enzymes. Part of the problem may be that too little attention is given to the subject in biochemistry courses, and the presentation is typically isolated and not reiterated throughout the course. This is understandable given the enormous content faculty feel they must deliver to students, a significant part of which is in service of the pre-medical curriculum. These problems might arise from a lack of coherency and repetition in the presentation of dynamic reactions. For example, it is hard to conceive that students could obtain an integrative understanding of the binding/chemical reactions and mathematical equations underlying enzyme kinetics without an adequate understanding of kinetics in their absence, which may not be described in textbooks or classes. The difficulty students have with dynamic reactions is consistent with the finding that they are the basis of key threshold concepts in biochemistry, the steady state and biochemical pathway dynamics and regulation[1].

Students also have difficulty expanding a sometimes-rote ability to analyze mathematical equations to a more intuitive understanding that bridges both mathematical and chemical equations. At the same time, depending on their training and department affiliation, instructors may not believe that they have a sufficient understanding of the necessary mathematics themselves, so they give it less attention than is needed for sufficient student success.

We believe that student understanding of dynamic biological reactions, which they mostly encounter in enzyme-catalyzed reactions, would ultimately be enhanced if they were given a stepwise introduction to progress curves, which show substrate [S] and product [P] concentrations with time. From a conceptual and intuitive perspective, this makes great sense. A substrate disappears and a product appears with increasing time. Each is associated with rate constants. In contrast, the initial velocity ($v_0$) used in Michaelis-Menten kinetic plots, is a derivative or secondary property much as density is a derivative property of mass and volume. As such, the initial velocity, as well as density (at least for young children), is less intuitive. Changes in substrate and product concentrations with time are exactly what students measure in the laboratory. Only with subsequent data analyses can the initial velocities $v_0$, the slope of [S] or [P] versus time curves, be calculated (as opposed to measured).

For any given progress curve model, mathematical equations must be derived, and statistics used to determine the closeness of the fit of the equation to actual data. Solving the equations for [S] or [P] versus time is always challenging, even for the simplest case, the conversion of a substrate to a product for a first-order irreversible reaction. The mathematical equation for the irreversible A → P reaction, which is typically taught in introductory chemistry classes, is shown in Equation 1 below.

$$Equation\ 1. \quad v = \frac{\Delta P}{\Delta t} = \frac{dP}{dt} = k_1 A$$

We will use the derivative of P vs t (the slope of the [P] vs t curve) as an expression of velocity, which is constantly changing during the reaction until equilibrium is achieved for reversible reactions or reactant concentration is zero for irreversible ones. For most introductory chemistry and biochemistry courses covering Michaelis-Menten kinetics, the initial velocity $v_0$ (determined at time t = 0) is most often used, even though it is difficult to determine accurately unless data at very early time points are available.

The equation $dP/dt = k_1A$ is an ordinary differential equation (**ODE**). Solving ODEs using integral calculus is beyond the training of most students taking biochemistry courses, and arguably instructors as well. Yet writing ODEs is not. Programs such as Copasi[2], Vcell[3,4], and Tellurium[5] are free and allow a mostly painless way to construct reaction diagrams and choose mathematical equations for each step of a reaction scheme. The computer then solves the ODEs numerically behind the curtain. For applications needed in a biochemistry class, two types of equations are generally selected, mass action and Michaelis-Menten type equations, although it is possible to write and solve simple Michaelis-Menten reactions using approximations of the more complicated Lambert-W Function, which give substrate and product concentrations as an explicit function of time[6].

The easiest and most intuitive to write are mass action equations that students studied in introductory chemistry courses.   Users can also select Michaelis-Menten irreversible or reversible equations, very familiar to those who have studied enzyme kinetics. In these programs, all mathematical equations are written as the flux (J) characterizing the conversion of a substrate to a product. Users can also input their own equations, such as for competitive, uncompetitive, and mixed inhibition reactions.   Copasi has a readily available list of 38 pre-defined equations.

Few references document the use of progress curves in undergraduate biochemistry courses [7–11].  At the same time, much has been written about the use of computer simulations and dynamic models to improve student learning [12–15]. Fundamentals of Biochemistry has two types of embedded interactive graphs.  One makes use of CalcPlot3D [16] which is used for simple binding reactions (saturation curves as a function of concentration) as well as for simple chemical reactions (progress curves) with easily derived integrated rate equations. Examples include irreversible and reversible first-order reactions such as A → P and A ↔ P.   For the other type of interactive graphs, we have developed and deployed a second simulation program, miniSidewinder, useful for both simple reactions and entire metabolic and signal transduction pathways.

For both types of graphs, sliders allow users to scale the graphs, change kinetic constants, and initial concentrations and see instantaneous updates in the progress curves. For the miniSidewinder software, simulations were constructed in Virtual Cell (Vcell) [3,4] or obtained from BioModels, a curated database for more extensive metabolic and signal transduction pathways [17]. (Programs like Copasi and Vcell can also be used to fit data using parameter scans, but these are not included in the book.) Vcell export files were used to create the interactive miniSidewinder graphs.   This program has an additional feature that allows users to export results to a spreadsheet file that shows concentration versus time for all reactants and products for a given set of constants used for the simulation.  Hence students can alter all constants to "interrogate" the models displayed in the book without having to create the simulation files.

For models in which the mathematical equations are accessible to students (for example those derived from mass action, Michaelis-Menten equations, and the Hill equation used for cooperative reactions or when a high-sensitivity response to an analyte is desired), the actual equations are presented in the book. The text contains fairly detailed explanations with some derivations for both non-catalyzed and enzyme-catalyzed chemical reactions as well.

Students should be able to write and understand individual ODEs for reactions, but they do not need to solve them, a task accomplished by the software.  Table 1 below shows examples of ODEs that students should be able to write (but not solve without available programs) using mass action principles.

| Reaction Scheme | Description | Example ODEs |
|---|---|---|
| A $\xrightarrow{k_1}$ P | 1st-order irreversible rx | $\frac{dA}{dt} = -k_1 A$  $\frac{dP}{dt} = +k_1 A$ |
| A $\underset{k_2}{\overset{k_1}{\rightleftarrows}}$ P | 1st-order reversible rx | $\frac{dA}{dt} = -k_1 A + k_2 P$ \| $\frac{dP}{dt} = k_1 A - k_2 P$ |
| A $\xrightarrow{k_1}$ B $\xrightarrow{k_2}$ C | Consecutive irreversible 1st-order rx | $\frac{dA}{dt} = -k_1 A$ \| $\frac{dB}{dt} = k_1 A - k_2 B$  $\frac{dC}{dt} = -k_2 B$ |
| A $\underset{k_{-1}}{\overset{k_1}{\rightleftarrows}}$ B $\underset{k_{-2}}{\overset{k_2}{\rightleftarrows}}$ C | Consecutive reversible 1st-order rx | $\frac{dB}{dt} = k_1 A + k_{-2} C - k_{-1} B - k_2 B$ |
| E + S $\underset{k_{-1}}{\overset{k_1}{\rightleftarrows}}$ ES | 2nd-order reversible binding E and S rx | $\frac{d[ES]}{dt} = k_1 [E][S] - k_{-1}[ES]$ |
| E + S $\underset{k_{-1}}{\overset{k_1}{\rightleftarrows}}$ ES $\xrightarrow{k_2}$ E + P | Irreversible enzyme-catalyzed S to P | $\frac{d[ES]}{dt} = k_1[E][S] - k_{-1}[ES] - k_2[ES]$ |

Table 1: Reaction schemes and their ordinary differential equations.

Each term on the right-hand side of the ODE has a rate constant that defines the formation/increase (+) or removal/decrease (-) of the species on the left-hand side.

**Mass Action vs Michaelis-Menten Equations: Which to use?**

Even though programs like Vcell and COPASI allow users to model and fit experimental data without integrating rate equations, it is still important for students to understand which model and set of equations are most appropriate to their needs. To illustrate this point we present in the book four different chemical models (ro, r1, r2, and r3), each with its mathematical formulation, for the irreversible conversion of S → P, as shown in Table 2 below. The first (r0) does not involve an enzyme and is simply the 1st order reaction conversion of substrate to product. The other three involve enzymes $E_n$. Reactions r0, r1, and r2 use mass action to write the ODEs for [P] vs time while reaction r3 uses the Michaelis-Menten equation. Of course, even this equation was derived from mass action and conservation principles.

| Rx/Description | Rx Diagram | Equations for $P_n$ |
|---|---|---|
| r0: Mass Action | Rx 0<br>S0 → P0<br>S $\xrightarrow{k_{f0}}$ P, $k_{r0}=0$ | $\dfrac{dP_0}{dt} = k_{f0}[S_0]$ |
| r1: Mass Action | Rx 1<br>E1, S1 → P1<br>E + S $\xrightarrow{k_{f1}}$ E + P, $k_{r1}=0$ | $\dfrac{dP_1}{dt} = k_{f1}[E][S_1]$ |
| r2: Mass Action | Rx 2<br>E2, S2 → ES2 (Rx 21) → P2 (Rx 22)<br>E + S $\underset{K_{r21}}{\overset{k_{f21}}{\rightleftarrows}}$ ES $\xrightarrow{k_{f22}}$ E + P, $k_{r22}=0$ | $\dfrac{dP_2}{dt} = k_{f22}[ES_2]$ |
| r3: Michaelis-Menten | Rx 3<br>S3 → P3<br>"$K_M$", $V_M$<br>E + S $\rightleftarrows$ ES → E + P | $\dfrac{dP_3}{dt} = \dfrac{V_M S_3}{K_M + S_3}$ |

Table 2: Comparison of mass action and Michaelis-Menten reactions and ODEs

Figure 1 below shows the progress curves for the formation of $P_n$ with time. The graph for $P_0$ is not shown for clarity since it is the same as that for $P_1$. This should be obvious from the mathematical equations shown in Table 2, since E1 is constant in reaction 2. The simple first-order exponential rise in P1 does not match those for the mostly superimposable curves for reaction 2 (more complex reaction model) and 3 (classical Michaelis-Menten model). This shows the importance of having students use a variety of models to fit data.

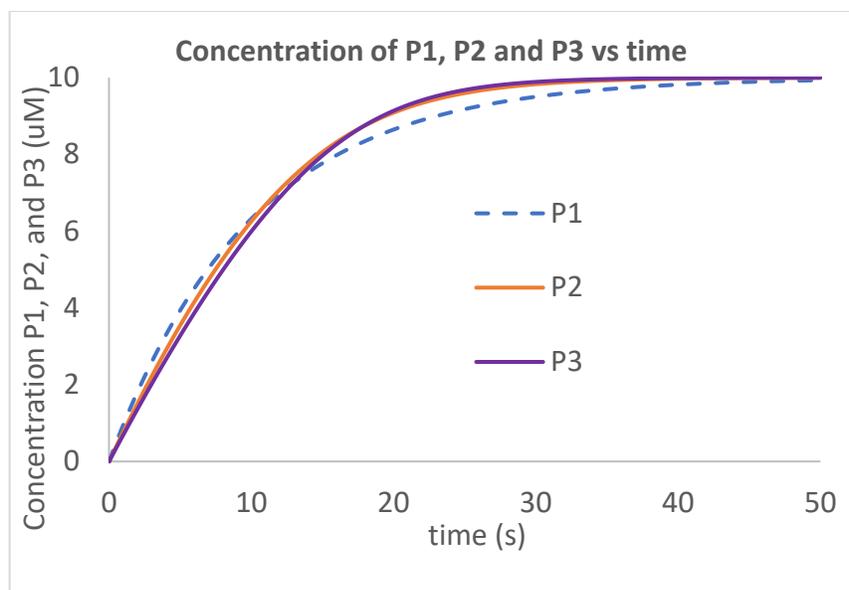

Figure 1: Progress curves for the formation of $P_n$ with time for mass action and Michaelis-Menten kinetic equations. P1 (- - -) is calculated from a simple mass action equation for reaction 1 (Table 2), while P2 (—) and P3 (—) are the mass action equations for reaction r2 and the Michaelis-Menten equation for reaction r3.

**Using the interactive graphs**

Dynamic graphs offer students the opportunity to actively interrogate a graph and hopefully develop a more fundamental understanding of the equations and kinetic constants. Here are two ways that students can interact with the graphs.

**a. Change rate and Michaelis-Menten constants**

The most important feature of the interactive graphs is to use the sliders to change the rate or Michaelis-Menten constants and immediately observe the effects on the concentration of reactants and products. The data for each set of rate constants can be exported as a csv file. Files exported for a variety of different runs can be used to create secondary plots. As an example, for the reversible reaction of A ↔ P, students could determine the dissociation constant $K_D = k_{rev}/k_{for} = 1/K_{EQ}$ at several different values of $k_{rev}$ and $k_{for}$ (first-order rate constants for the reverse and forward reactions, respectively) to help them understand that thermodynamic equilibrium and dissociation constants are in effect the ratios of rate constants, giving them a more intuitive understanding of the relationship between rate and equilibrium/dissociation constants. Another important simulation in the book allows students to change the rate constants for an isolated reversible enzyme and compare its kinetics to an enzyme with the same kinetic constants embedded in a "mini-pathway" with an upstream constant input of substrate and downstream removal of product. This would illustrate that the calculated $K_D$ for the enzyme is in effect a steady-state $K_D$ which is not equal to the thermodynamic $K_D$ for the isolated enzyme. Indeed, differentiating the steady state in an open biological system from the equilibrium state in a closed system is a key threshold concept[1] whose understanding eludes most students.

### b. Change concentrations

CSV files derived from time course simulations conducted at a variety of substrate or inhibitor concentrations could be used to produce secondary plots when data for all of the substrates are on one graph. The progress curve data (dP/dt) could be used to calculate the initial velocity $v_0$ for each substrate/inhibitor concentration. Students could then make secondary plots of $v_0$ vs [S] and determine the Michaelis-Menten constants $V_M$ and $K_M$ by fitting the data to the Michaelis-Menten equation using nonlinear regression analyses, or by making Lineweaver-Burk plots of $1/v_0$ vs $1/[S]$ and fitting them with a weighted linear regression model. This would give students yet another way to bridge their understanding of time course data (which again is what students measure in the lab) and initial velocity calculations and their replots to obtain kinetic parameters.

The data exported from the simulations are "perfect" without noise. To give students more realistic data for their secondary $v_0$ vs [S] replots, random "noise" could be added to [P] values in a spreadsheet by multiplying the [P] values by 0.5*(RAND()-0.5) where the first 0.5 is a scale factor. This formula produces a noise term of constant magnitude with the scale factor determining the magnitude of the noise term.

### Simulation of complex reaction schemes and pathways

Mathematical and computer modeling, and increasingly machine learning and AI, are critically essential if we wish to understand the complexity of life. Yet until recently (until the need to analyze and compare vast amounts of structural and metabolic data), biology has not been considered a mathematical discipline. Intuition is no longer (as if it were ever) sufficient to even qualitatively predict the inputs and outputs of interconnected systems. Indeed as with the steady state, biochemical pathway dynamics and their regulation also represent a threshold concept[1].

One clear and very simple example where intuition fails occurs when the concentrations of species change cyclically with time. Some may have encountered the oscillating Briggs–Rauscher reaction in introductory chemistry courses in which iodate ($IO_3^-$, colorless) is oxidized to $I_2$ [18]. $I_2$ then reacts with malonic acid to produce $I^-$ (colorless). This can react in a rapid step with $I_2$ to produce $I_3^-$, which binds to starch to form a blue color. With time the $I_3^-$ is reconverted back to $I_2$ and $I^-$ and the reactions can repeat. The reactions are summarized in Equations 1-4 below [18].

$$2H^+ + 2IO_3^- + 5H_2O_2 \xrightarrow{Mn^{2+}} I_2 + 5O_2 + 6H_2O \quad (1)$$

$$CH_2(COOH)_2 + I_2 \rightarrow CHI(COOH)_2 + I^- + H^+ \quad (2)$$

$$I_2 + I^- \rightarrow I_3^- + starch \rightarrow I_3^- - starch\ complex \quad (3)$$

$$I_3^- - starch \rightarrow\rightarrow I_2 + I^- \quad (4)$$

Although useful for a demonstration, the chemistry of this modified iodine clock reaction is likely less engaging to biochemistry students than the following examples of a cyclic reaction in signaling pathways.

### Mitogen-Activated Protein Kinase (MAPK) Activation with Oscillatory Negative Feedback

Figure 2 shows the activation of MAPK by two upstream protein kinases that act in succession. The first, MAP kinase kinase kinase (MKKK) is activated by phosphorylation by a kinase not shown. The active

MKKK_P then doubly phosphorylates the next downstream protein kinase MKK to MKK_PP. This then activates MAPK by doubly phosphorylating it to MAPK_PP. Students would rightly guess that a progress curve of d[MAPK_PP]/dt would go from zero and plateau with time. However, no one, without a computational simulation, would intuit that [MAPK_PP] would oscillate like the iodine clock reaction if the active MAPK_PP, in a feedback "product" inhibition step, inhibits the activating phosphorylation of the MKKK, the first enzyme in this pathway [19]. The Vcell reaction diagram for the MAPK activation cascade is shown in Figure 2.

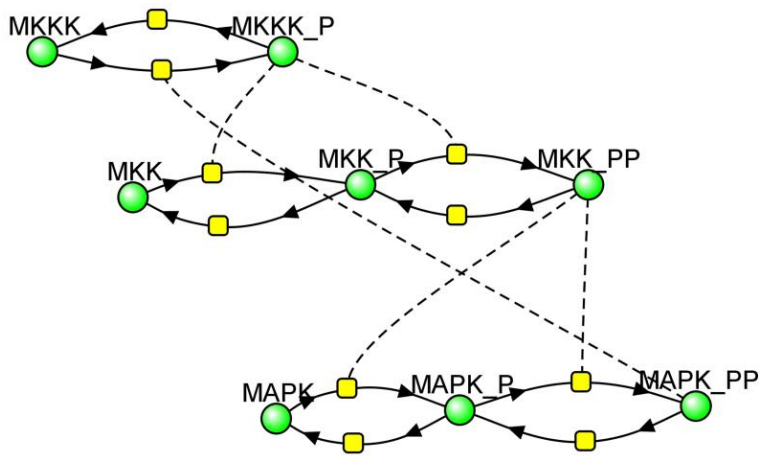

Figure 2: Reaction scheme for the cascade activation of MAPK

The progress curves for d[MAPK_PP]/dt for both the uninhibited and feedback-inhibited reactions are shown in Figure 3.

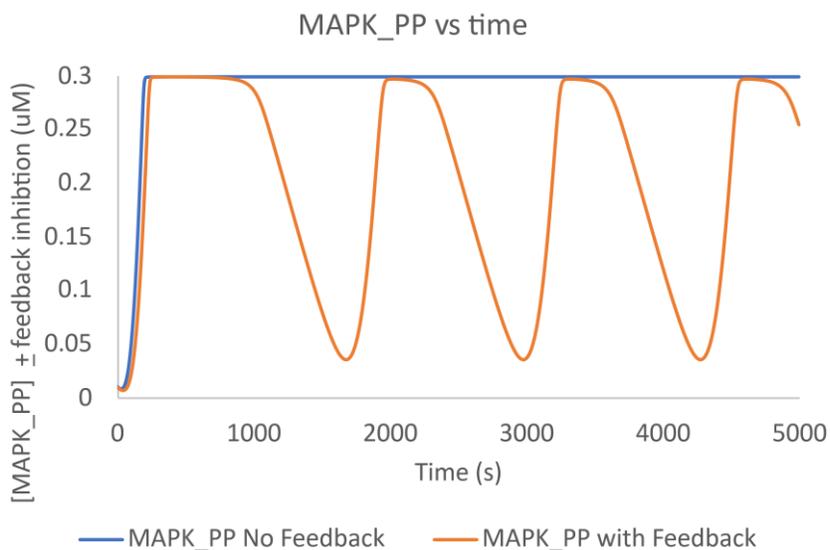

Figure 3: Comparison of the progress curves of d[MAPK_PP] vs time in the absence and presence of feedback phosphorylation of MKKK

Students can use the appropriate slider to reduce or turn off the feedback inhibition step and immediately observe the effects on the oscillations. Since this cascade is catalytic, each step amplifies the other, which makes the pathway ultrasensitive to changes in the concentrations and rate constants. Such oscillations would then propagate spatially in the cell.

**Perfect Adaptation:  Negative Feedback Loop**

Equally non-intuitive mathematically but elegant in their simplicity are small 2-3 protein reaction motifs that display perfect (or near perfect) adaptation [20], a necessary condition of homeostasis.  In perfect adaptation, an output returns to a basal or near basal state, even in the presence of increasing stimuli. An example of a common motif, the negative feedback loop, is shown in Figure 4.

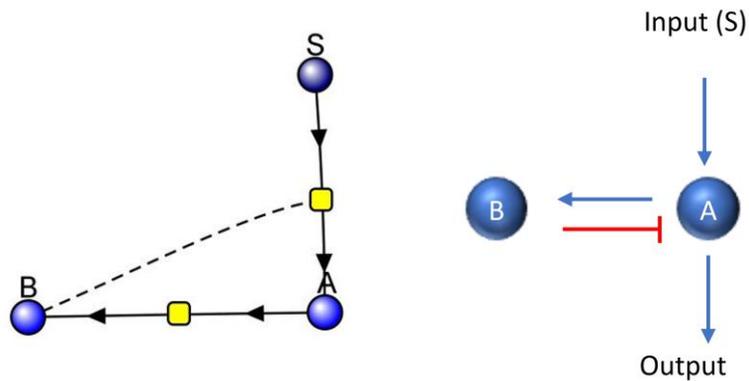

Figure 4:  Reaction scheme for perfect adaptation by a negative feedback loop

A graph showing a stepwise stimulus S, with negative feedback inhibition by a product B made from A, and the output A, is shown in Figure 5.

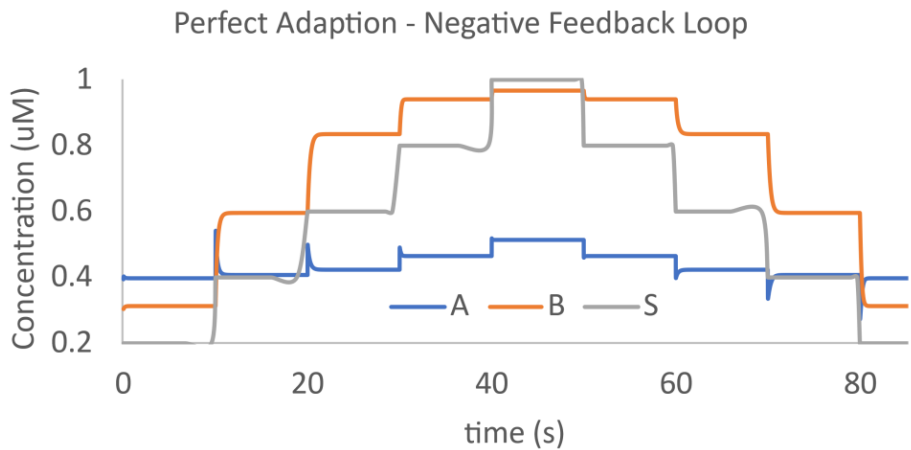

Figure 5:  Progress curves for perfect adaption by a negative feedback loop with a step function increase and decrease in the concentration of a stimuli S

The blue line is the response, designated in this model as A.  B acts as an inhibitor (note the dotted line to the input node in the left diagram and the blunt-ended red bar in the middle diagram.  The stimulus goes from 0.2 uM (initial concentration) at t =0  to 1 uM (a 5-fold increase) at 40 seconds, but the response A increases at most from 0.4 (initial condition) to 0.5 (a 1.25-fold increase).

Present versions of the simulation software in the book can <u>not yet run this model</u> which has a special "event protocol" to create the stepwise increase and decrease in the stimuli. In addition, the Vcell file can not be exported as a systems biology markup (sbml) file necessary for simulation in the book. Hence the present version of the book just shows the reaction scheme with static graphs.

**Reaction Pathways**

Entire reaction pathways can be simulated in the book using miniSidewinder. Two examples include anaerobic [21] and aerobic glycolysis in yeast. The time course of some key species in yeast anaerobic glycolysis is shown in Figure 6

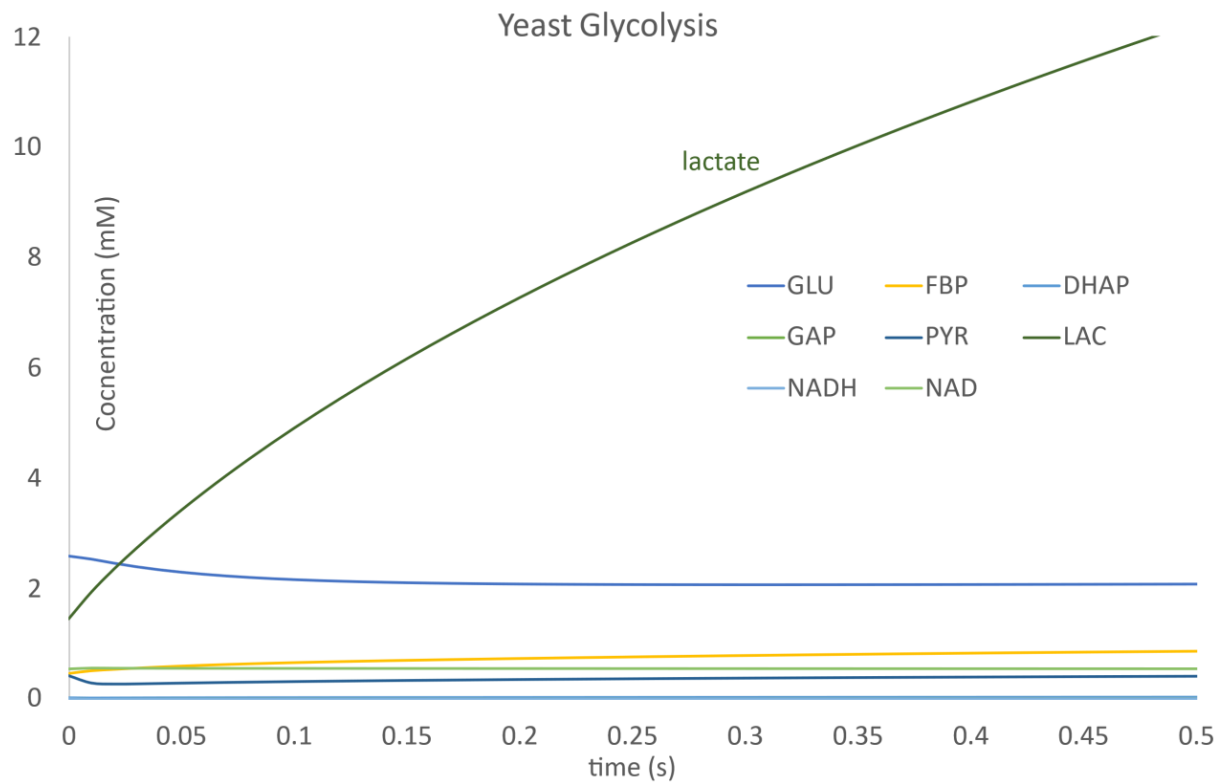

Figure 6: Progress curves for key species in yeast anaerobic glycolysis.

Students can interrogate the live graphs by changing rate constants and concentrations and see their effects on flux at each step in the pathway.

**Simulation Program: Development and Use**

The main interactive computational simulation program used in the Fundamentals of Biochemistry OER, miniSidewinder, is an open-source, browser-based, client-side application (https://github.com/sys-bio/miniSidewinder). It works on any computer with a web browser (Microsoft Edge, Google Chrome, Mozilla Firefox, Apple Safari, etc) and has minimal server-side requirements. The simulator makes use of the Systems Biology Markup Language (SBML) standard for describing mathematical/chemical network models (https://sbml.org/). The SBML data format allows model interoperability between any simulation software that supports SBML, enabling easy model sharing and modification [22].

**Usage**

For the website author, embedding the simulator and model into an educational webpage involves:

1. Uploading the associated simulator HTML and JavaScript files to the website;
2. Uploading the SBML formatted model to the website;
3. Inserting an HTML object into the website page that contains a javascript function template specifying the model, simulation parameters, and location of the simulator.

This approach is used by the Fundamentals of Biochemistry and a Google Sites page for interactive models: https://sites.google.com/view/interactive-modeling/ .

Website model users can:

1. Change sliders to adjust the values of parameters and species. After each adjustment, a simulation is run and plotted. Adjusting these values produces immediate feedback on the effect these changes have on the simulation giving the user a better understanding of the model's behavior.
2. Select buttons to change the parameter and species sliders, plot species, and y plot scaling. A log scale is currently unsupported.
3. Download a comma-separated values (.csv) file of the last simulation run for further analysis.
4. Change the simulation run time and the integrator time step size. This gives the user the ability to investigate the behavior of the model at different time scales.

From the example mentioned earlier, MAPK activation with oscillatory feedback, Figures 7 and 8 show how the user can interact with the simulator to explore model behavior. At the bottom of each plot are sliders labeled with parameters and species. By adjusting one of the parameter sliders the parameter's effect on the system can be observed. Figure 7 shows the MAPK oscillatory feedback but Figure 8 shows that when the inhibitory constant Ki_J0 is increased, the oscillation ceases. In this way, the user can investigate the effects of each parameter on the system gaining more insight into the model than could be done by just inspecting the model equations or reaction diagrams.

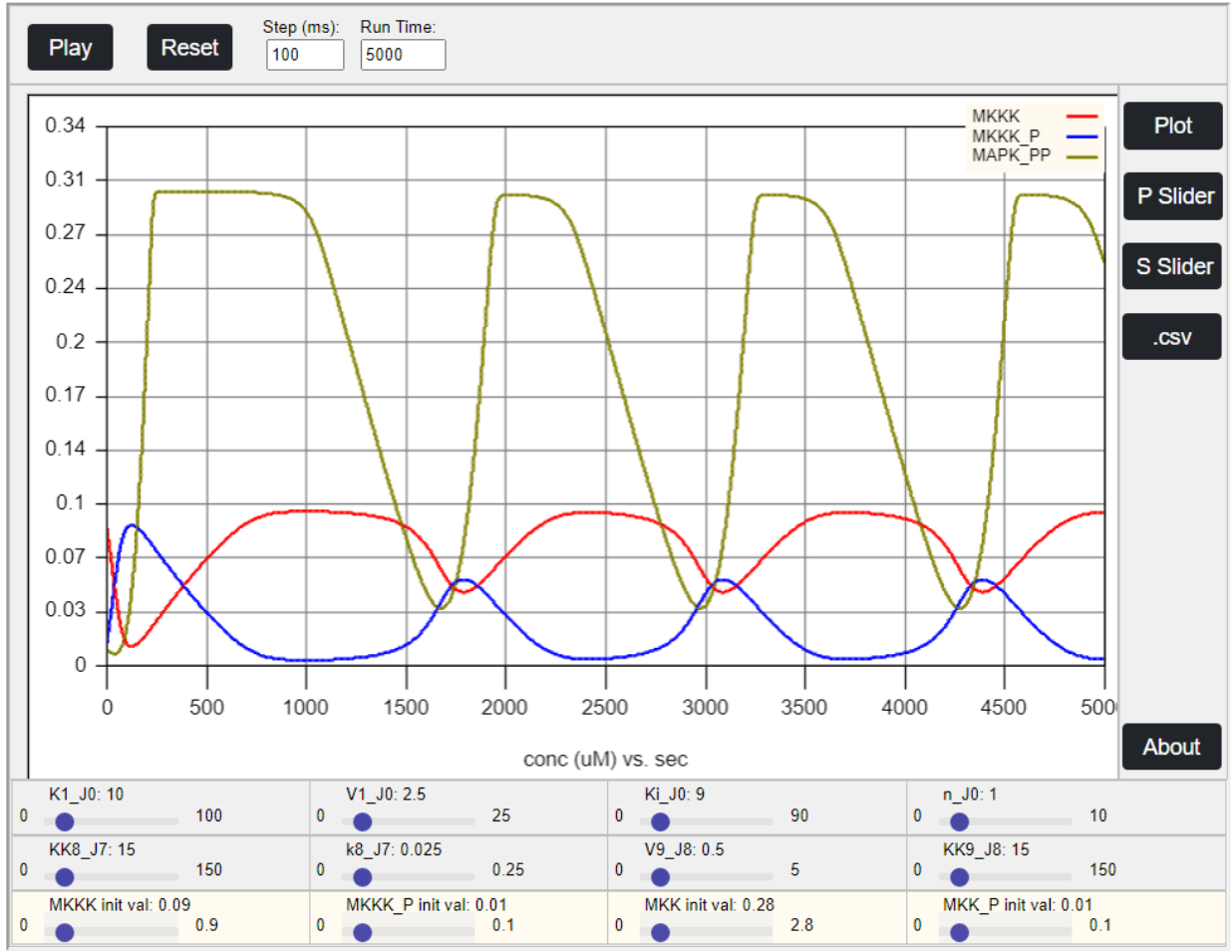

Figure 7: Simulation plot of MAPK activation with oscillatory feedback. MAPK_PP [olive] oscillates. Note parameter value and species initial condition sliders at the bottom. As a user moves a slider the simulation is repeated with the new slider value.

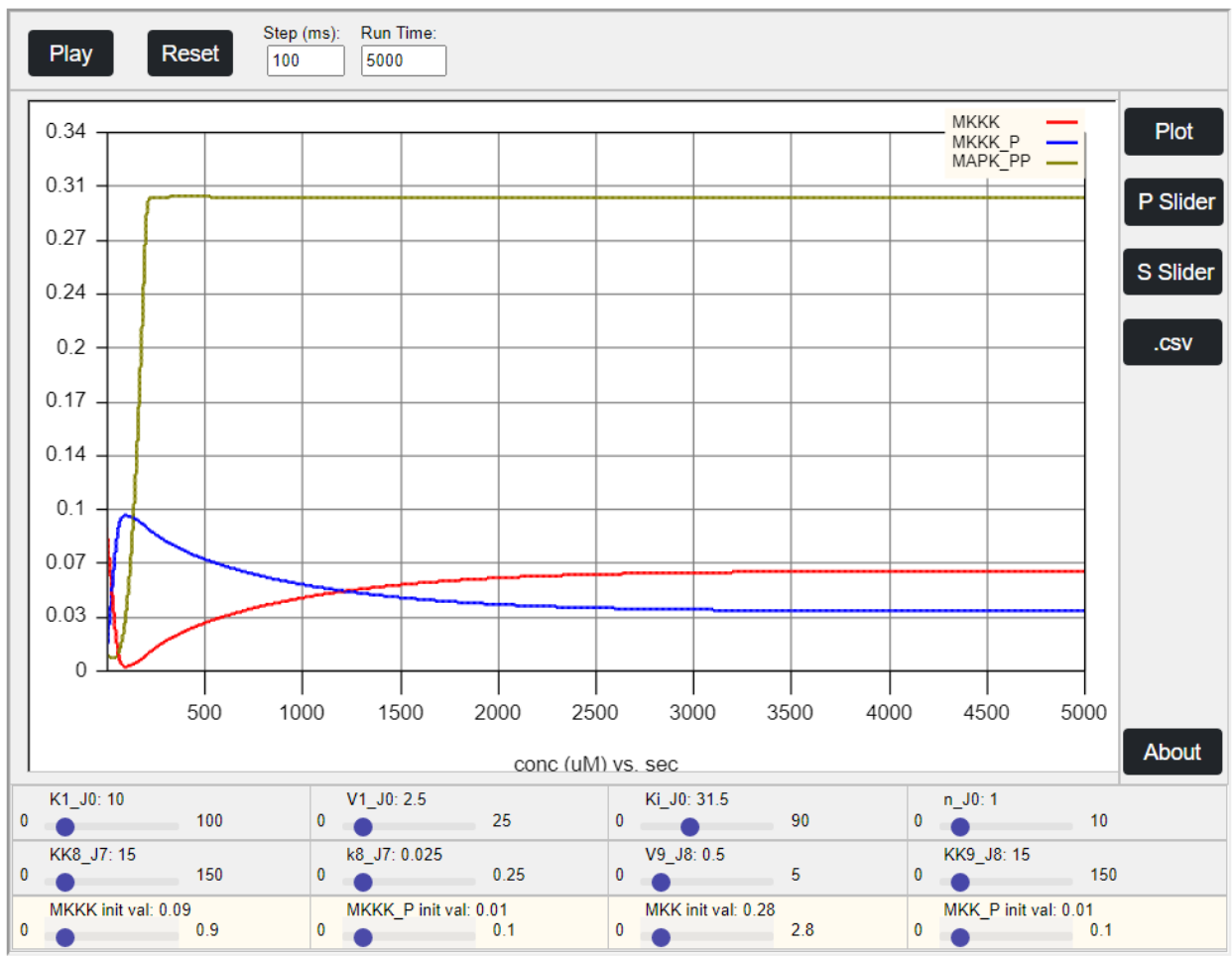

Figure 8. MAPK activation with no oscillatory feedback, MAPK_PP [olive] no longer oscillates. By changing the slider 'Ki_J0', corresponding to an inhibitory constant for the MKKK -> MKKK_P flux J0, from a value of '9' to '31.5' nmol/l, the oscillatory behavior of the progress curves disappears giving insight into how the system of reactions behaves when the inhibitory constant changes value.

MiniSidewinder is deployed throughout the Fundamentals of Biochemistry, although there are a few rare cases where a "bulkier" simulation program is utilized. This alternative simulation program relies on BioSimulatiors[23], which submits jobs to an external server and thus is able to run simulations using the COMBINE archive (omex) file format. However, this reliance on an external server makes BioSimulators noticeably slower than the near-instantaneous responsiveness of miniSidewinder and thus makes it more suitable for the textbook in all but these rare exceptions. All models using the miniSideWinder[24] and Combine archive (omex)[25] software deployed in Fundamental of Biochemistry are available in the book.

**Conclusion**

We have incorporated computational simulation software into Fundamentals of Biochemistry which allows users to change rate constants and concentrations of all species and instantaneously observe changes in the progress curves for all species. The book includes simulations that vary in complexity from a single first-order reaction to full metabolic pathways.  The next step in the deployment of these simulations is to create and embed assessment questions into the text.  However, faculty who use the text can easily create questions specific to their use and needs.

At present, we have no assessment data that would measure the effect of the use of the simulations on student learning.  We welcome faculty interested in assessment to do that.  Likewise, we welcome ideas for new simulations and will gladly work with interested faculty to incorporate them into the book.

Simulation Engines and Services for Recommending Specific Tools. *Nucleic Acids Res* **2022**, *50* (W1), W108–W114. https://doi.org/10.1093/nar/gkac331.

(24) Jakubowski, H. *Sidewinder Models*. https://bio.libretexts.org/Learning_Objects/Visualizations_and_Simulations/Progress_Curve_Analysis/SBML_Computational_Models (accessed 2023-06-27).

(25) Jakubowski, H. *COMBINE Archive Models*. https://bio.libretexts.org/Learning_Objects/Visualizations_and_Simulations/Progress_Curve_Analysis/VCell_-_Computational_Models (accessed 2023-06-27).
**Acknowledgements**

Funding:

*MiniSidewinder*: NIH/NIGMS (Grant R01-GM123032-04)

*LibreText*: Department of Education Open Textbook Pilot Project, the UC Davis Office of the Provost, the UC Davis Library, the California State University Affordable Learning Solutions Program, and Merlot.
17